\documentclass[twocolumn,prl,nofootinbib]{revtex4-1}
\usepackage{graphicx}
\usepackage{amssymb}
\usepackage{epstopdf}
\usepackage{amsmath,bm}
\usepackage{footnote}
\usepackage[bottom]{footmisc}
\usepackage{color}
\usepackage{physics}
\makeatletter

\@ifundefined{textcolor}{}
{
 \definecolor{BLACK}{gray}{0}
 \definecolor{WHITE}{gray}{1}
 \definecolor{RED}{rgb}{1,0,0}
 \definecolor{GREEN}{rgb}{0,1,0}
 \definecolor{BLUE}{rgb}{0,0,1}
 \definecolor{CYAN}{cmyk}{1,0,0,0}
 \definecolor{MAGENTA}{cmyk}{0,1,0,0}
 \definecolor{YELLOW}{cmyk}{0,0,1,0}
}

\usepackage[colorlinks,
            linkcolor=black,
            anchorcolor=black,
            citecolor=black
            ]{hyperref}
\usepackage{ulem}           
            
\makeatother

\begin{document}

\title{Magnetic Topological Kagome Systems}

\author{Julian Legendre$^{1}$ and Karyn Le Hur$^{1}$}
\affiliation{$^1$  CPHT, CNRS,  Institut Polytechnique de Paris, Route de Saclay, 91128 Palaiseau, France} 
\date{\today}

\begin{abstract}
The recently discovered material Co$_3$Sn$_2$S$_2$ shows an impressive behavior of the quantum anomalous Hall (QAH) conductivity driven by the interplay between ferromagnetism in the $z$ direction and antiferromagnetism in the $xy$ plane. Motivated by these facts, first we build and study a spin-1/2 model to describe the magnetism of Co-atoms on the Kagome planes. Then, we include conduction electrons which are coupled to the spins-1/2 through a strong Hund's coupling. The spin-orbit coupling results in topological low-energy bands. For 2/3 on-site occupancy, we find a topological transition from a QAH ferromagnetic insulating phase with Chern number one to a quantum spin Hall (QSH) antiferromagnetic phase. The QAH phase is metallic when slightly changing the on-site occupancy. To account for temperature effects, we include fluctuations in the direction of the Hund's coupling. We show how the Hall conductivity can now smoothly evolve when spins develop a $120^o$ antiferromagnetism in the $xy$ plane and can synchronize with the ferromagnetic fraction.
\end{abstract}
\maketitle

{\it Introduction.---} When applying a magnetic field, the quantum Hall effect gives rise to an insulating  behavior in the bulk of a material and is characterized by chiral edge states \cite{QHE,Halperin,Buttiker} which show a quantized Hall conductance. Bulk properties are described through a topological invariant, the Chern number \cite{TKKN}. The QAH effect, as originally introduced by Haldane \cite{Haldane2D}, corresponds to a generalization of the quantum Hall effect on the honeycomb lattice with tunable Berry phases. It opens a gap for the Dirac fermions and breaks time-reversal symmetry, such that a unit cell yet shows a zero net flux. This model finds applications in quantum materials \cite{ReviewMaterial,Cavalleri}, light \cite{ReviewQED,Alberto} and cold atom systems \cite{Jotzu,bilayer}, and was developed in other geometries such as the Kagome lattice \cite{Kagomelight,KagomeNagaosa}. For practical realizations, it is important to find intrinsic ferromagnetic QAH systems with topologically non-trivial band gaps produced by spin-orbit coupling mechanisms \cite{ZhangSC}. The weyl-semimetal quantum material Co$_3$Sn$_2$S$_2$ has recently attracted a lot of attention experimentally in relation with the QAH effect \cite{Princeton,Dresden}. The pure cobalt is known to have a Curie temperature of around 1388K associated to ferromagnetism. Here, a layered crystal structure with a Co-Kagome lattice in this material develops a perfectly out-of-plane ferromagnetic phase (along $z$ direction) and an almost quantized Hall conductivity under 90K. Between 90K and 175K, the ferromagnetic fraction smoothly decreases while an in-plane antiferromagnetism (related to the $xy$ plane) progressively develops \cite{Princeton}. The anomalous Hall conductivity then evolves with the ferromagnetic fraction along $z$ direction \cite{Princeton,Dresden}. Inspired by these realizations, we introduce a new class of magnetic topological Kagome systems. 

\begin{figure}[t]
   \includegraphics[width=7.5cm]{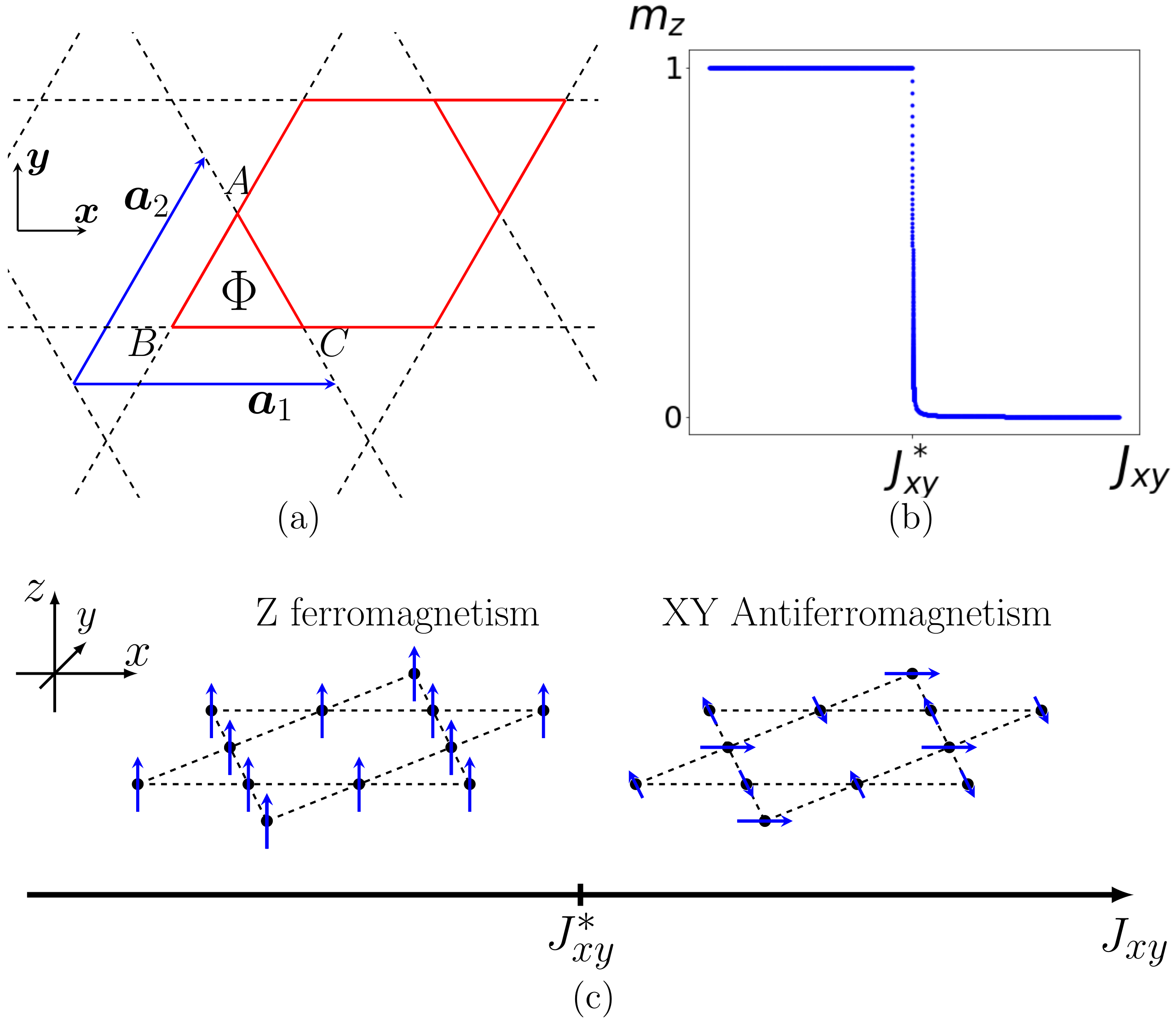}
   \caption{(a) Kagome lattice. Here, $\Phi$ is the phase accumulated by an electron hopping counter clockwise inside a triangle. ${\bf a}_1$ and ${\bf a}_2$ are the two vectors characterizing a unit cell and their norm is set to unity, $|{\bf a}_1|=|{\bf a}_2|=1$. (b) Magnetization for the core spins $ m_z =\cos\theta$ along the $z$ axis when neglecting fluctuations in the direction of the Hund's coupling, for a (very) small magnetic field $h_a$. (c) Sketch of the magnetic transition when increasing the $XY$ coupling if we fix $J_z=1$.}
          \label{sketch}
    \vskip -0.5cm
\end{figure}

Within our approach, the magnetism of Co-atoms is described through localized spins \cite{Khaliullin}, reflecting the strong Hubbard interaction, and the low-energy bands are in agreement with ab-initio calculations on Co$_3$Sn$_2$S$_2$ established in the ferromagnetic phase \cite{Princeton,shandite,Nature}. The magnetic transition is described through the localized spins and itinerant electrons will develop topological energy bands, as a result of the spin-orbit coupling.
While Kondo lattices have been shown to induce topological phases \cite{Dzero}, here itinerant and localized electrons (the latter forming core spins-1/2 on each atom) are coupled through a strong Hund's ferromagnetic mechanism, as also suggested in Ref. \cite{Rosales}. The presence of a Hund's coupling generally plays a key role in these multi-orbital electronic systems \cite{Rice,KagomeNagaosa}. In our model, this coupling is along the $z$ direction. It induces an Ising $J_z$ ferromagnetic interaction between nearest-neighbor localized spins \cite{Zener,Anderson,Jonker,PGdeGennes}, reproducing the ferromagnetism of the Co-atoms below 90K \cite{Princeton}. 
We also introduce an in-plane antiferromagnetic correlation $J_{xy}$ between the core spins which is produced by Mott physics and electron-mediated interactions between the half-filled orbitals (associated to the localized spins) \cite{Karyn}. This model produces an antiferromagnetic transition with a $120^o$ spin ordering in the $xy$ plane when $J_{xy}^*=2J_z$ (see Fig. \ref{sketch}), as observed \cite{Princeton}.

From a spin-wave analysis, a flat band touches the classical ferromagnetic state when approaching the transition, destabilizing the ferromagnetic alignement and stabilizing the antiferromagnetic $120^o$ spin ordering in the $xy$ plane. The flat band then moves to higher energy because the azimuthal angle $\phi_i$ associated to each spin on the Bloch sphere is now only a global quantity, since we fix $\phi_i-\phi_j=2\pi/3$ in radians (or $120^o$), and the polar angle of each spin jumps to $\theta_i=\pi/2$. The magnetization along $z$ direction jumps discontinuously to zero. It becomes continuous if we apply a small magnetic field. Then, we describe temperature effects in Co$_3$Sn$_2$S$_2$ below $175K$ by decreasing the ferromagnetic $J_z$ coupling or equivalently by increasing the antiferromagnetic coupling $J_{xy}$ if we set $J_z=1$.  Taking into account fluctuations in the direction of the Hund's coupling produces a (Gaussian) distribution on the value of $J_z$. At finite temperatures, the formation of magnetic domain walls, as recently observed with imagery analysis \cite{domains}, could also justify this statistical view. Interestingly, we then find that the (average) system's magnetization in $z$ direction smoothly reduces to zero after the transition producing the progressive canting of the spins, such that the statistically averaged Chern number follows the ferromagnetic fraction. 

{\it Model.---} The mechanism leading to the anomalous Hall effect in our model is the intrinsic spin-orbit coupling (SOC), which may originate from the presence of Sn$_2$ atoms \cite{Ozawa}. Kane and Mele showed that the SOC can produce
a QSH phase on the honeycomb lattice \cite{KM}. This phase (called a $\mathbb{Z}_2$ topological insulator) is characterized by spin-up and spin-down electrons at the edge moving in
opposite direction, producing a vanishing Hall conductance.  A QSH effect was also predicted and observed in two-dimensional Mercury \cite{Bernevig,HgTe} and in three-dimensional Bismuth \cite{ReviewQSH} quantum materials. In the Kane-Mele model, strong interaction effects in the Mott phase favor an in-plane antiferromagnetic phase \cite{topoMott}, justifying that we choose an antiferromagnetic $XY$ spin coupling for the core spins in addition to $J_z$. A link between SOC and QAH effect was also studied for Cs$_2$LiMn$_3$F$_{12}$ \cite{ZhangSC} and in relation with chiral spin states \cite{KagomeNagaosa,Rosales}. 

Below, we include the effect of the competition between the two magnetic channels $J_{xy}$ and $J_z$, onto the probabilities of occupancies $P_{\uparrow}$ and $P_{\downarrow}$ for the spin up and spin down itinerant electrons in the canonical ensemble, assuming a strong Hund's coupling:
\begin{equation}
\label{hc}
H_c = -\frac{h_c}{2}\sum_i S_i^z(c^{\dagger}_{i\uparrow}c_{i\uparrow}-c^{\dagger}_{i\downarrow}c_{i\downarrow}),
\end{equation}
where $S_i^z$ refers to the z-magnetization of the localized spin-1/2 at site $i$ and $c^{\dagger}_{i\alpha}$ creates a conduction electron at site $i$ with spin polarization $\alpha=\uparrow,\downarrow$. The $J_z$ spin coupling is induced by the Hund's coupling $h_c$. 

\begin{figure}[t]
   \includegraphics[width=7.5cm]{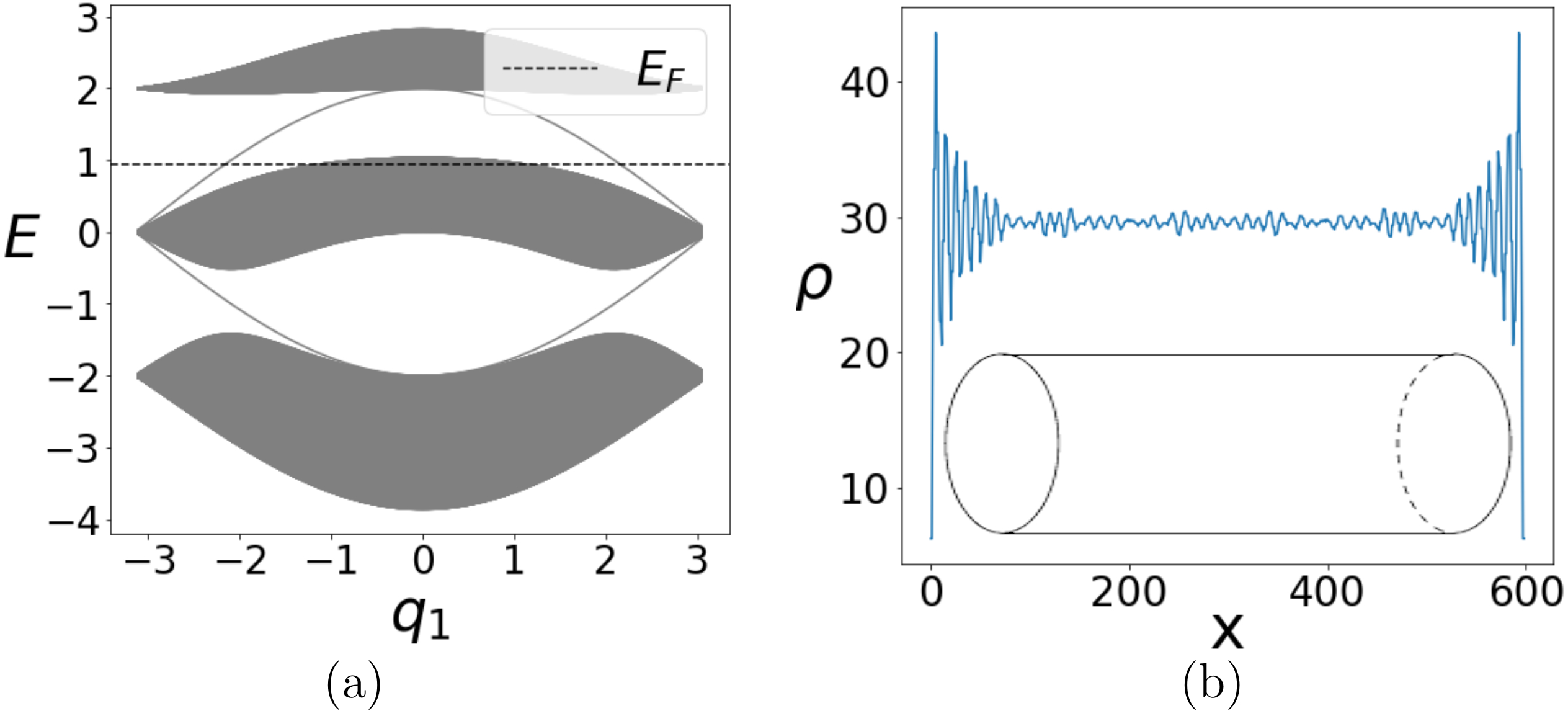}
   \caption{(a) Energy relation dispersion relation and topological edge modes computed for a lattice cylinder geometry. Vectors ${\bf q}_1$ and ${\bf q}_2$ are dual to vectors ${\bf a}_1$ and ${\bf a}_2$ in Fig. 1.  (b) Local density of states $\rho$ for a Fermi energy at $E_F=0.95$ (in units of $t$) with associated cylinder geometry and with flux inside each triangle of the lattice $\Phi=3\pi/4$. Here, $x$ refers to the number of Co-atoms along the cylinder direction.}
          \label{energyband}
    \vskip -0.5cm
\end{figure}

We address the case where the on-site occupancy for itinerant electrons is 2/3 and close to 2/3. If the system is spin-polarized with one electron species, when the Fermi energy lies in the gap between the middle and the upper energy bands in Fig. \ref{energyband}, the system will show a quantized Hall response. Changing slightly the on-site occupancy it is also possible to observe a metallic ferromagnetic topological phase \cite{Petrescu,Haldane2004}. To tackle the ferromagnetic-antiferromagnetic transition, we may introduce the number of particles associated to up and down species, as  $N_{\uparrow}=P_{\uparrow}N_e$ and $N_{\downarrow}=P_{\downarrow}N_e$ with the number of electrons $N_e=N_{\uparrow}+N_{\downarrow}$ satisfying $N_e=\frac{2}{3}N_a$ and $N_a$ being the number of atomic sites. When the number of up and down electrons is equal, then the lowest energy band associated to each spin species is filled. 

When $J_{xy}\ll J_z$, the spins of the electrons adiabatically follow the polarization of the core spins due to the strong $h_c$ coupling, such that $c_{i\uparrow}=c_i$, and one can build an effective spin polarized electron model with only spin-up electrons. The tight-binding model for the spin-up electrons takes the form: $H_{QAH} = \sum_{\langle i; j\rangle} -(t+i\nu_{ij}\lambda)c^{\dagger}_i c_j$, where $t$ (real) is the nearest-neighbor hopping amplitude on the Kagome lattice and $\lambda$ the intrinsic spin-orbit coupling projected onto the spin-up electronic states \cite{ZhangSC}. Here, $\nu_{ij}=+1(-1)$ if the electron jumps counter-clockwise (clockwise) inside the triangle of the Kagome lattice containing sites $i$ and $j$, and the symbol $\langle i;j\rangle$ refers to a coupling between nearest neighbors. We observe that the ferromagnetism should not modify the hopping amplitude of spin-up electrons compared to the case where $\langle S_i^z\rangle=0$, implying that $t_{\uparrow}=t=t\langle\chi_i | \chi_j\rangle$, with $|\chi_i\rangle$ representing a spin eigenstate at site $i$ with $\theta_i=0$.

To make a link with the Haldane model on the Kagome lattice, we can then re-write $-(t+i\nu_{ij}\lambda)=-re^{i\Phi\nu_{ij}/3}$ with $r=\sqrt{t^2+\lambda^2}$ and $\Phi=3\hbox{arg}(t+i\lambda)$. In Fig. \ref{sketch}, in a triangle there is a flux $\Phi$ breaking time-reversal symmetry and in an hexagon (honeycomb cell) there is a flux $-2\Phi$ such that globally on a parallelogram unit cell represented by the vectors ${\bf a}_1$ and ${\bf a}_2$ the total net flux is zero. In wave-vector space, for 2/3 on-site occupancy, we then check the presence of a QAH effect for an illustrative value of $\Phi=3\pi/4$, see Fig. \ref{energyband}(a); see Supplemental Material in Ref. \cite{SM} for methodology. The 3 energy bands reflect the three distinct sites $A$, $B$, $C$ in Fig. \ref{sketch}. The lowest energy band is described by a Chern number $C_l=\hbox{sgn}(\sin\Phi)=+1$, the middle band has a total Chern number zero, and the upper band shows a Chern number $C_u=-C_l$. It is important to remind that the middle band becomes perfectly flat for $\Phi=\pi/2$ and it touches the bottom of the lowest band for $\Phi=0$, suppressing the QAH effect when $\lambda=0$. In Fig. \ref{energyband}(b), the local density of states for on-site occupancy close to 2/3 shows a ferromagnetic topological phase with a metallic bulk  and with a Chern number almost equal to $1$, as observed \cite{Princeton,Nature}.

{\it Magnetic Transition.---} Now, we study quantitatively the magnetic properties of the system in the presence of the couplings $J_z$ and $J_{xy}$. The localized spins are described by the Hamiltonian:
\begin{equation}
H_{S} = \sum_{\langle i;j\rangle} [-J_z S_i^z S_j^z + J_{xy}(S_i^xS_j^x + S_i^y S_j^y)],
\end{equation}
with $(J_z,J_{xy})>0$, such that the classical energy on the Bloch sphere representation is:
\begin{equation}
E = \frac{1}{4}\sum_{\langle i;j\rangle}[-J_z\cos\theta_i\cos\theta_j + J_{xy}\sin\theta_i\sin\theta_j\cos(\phi_i-\phi_j)].
\label{bloch}
\end{equation}
To minimize the magnetic energy, we find that $\theta_i=\theta$ for all values of $J_{xy}/J_z$ and in the antiferromagnetic phase $\phi_i-\phi_j=2\pi/3$. Therefore, the classical energy takes the simple form $E=\frac{1}{4} \sum_{\langle i;j\rangle} [(-J_z+\frac{1}{2}J_{xy})\cos^2\theta - \frac{J_{xy}}{2}]$. For $2J_z> J_{xy}$ the energy reaches its minimum for $\theta=0$ or $\theta=\pi$, corresponding to $E=-N_a J_z$ and to a ferromagnetic state of the spins along $z$ direction.  For $2J_z=J_{xy}$, the ground state energy takes the value $E=-\frac{N_a}{2} J_{xy}$ for all the values of $\theta$. For $J_{xy}>2J_z$, the ground state energy keeps the same value $-\frac{N_a}{2} J_{xy}$ if the spins now point in the $xy$ plane with $\theta_i=\theta=\pi/2$.
Then, we study the effect of a small applied magnetic field $h_a$ along $z$ direction which favors the classical minimum $\theta=0$ when $2J_z+h_a\geq J_{xy}$, corresponding to an energy $E(\theta=0)=-N_a(J_z+h_a)$. If $J_{xy}>2J_z+h_a$, then $E$ is minimum for $\theta$ such that
\begin{equation}
\label{costheta}
\cos\theta=\frac{h_a}{-2J_z+J_{xy}},
\end{equation}
resulting in $E = -\frac{N_a}{2}\left(\frac{h_a^2}{-2J_z+J_{xy}} +J_{xy}\right)$. This behavior associated to the magnetization along $z$ direction is shown in Fig. 1 (top right). While the SU(2) Heisenberg antiferromagnetic Hamiltonian shows a quantum spin liquid on the Kagome lattice \cite{Kagomeliquid}, here magnetic ordered phases are stable classically through the form of $H_S$. 

\begin{figure}[t]
   \includegraphics[width=9cm]{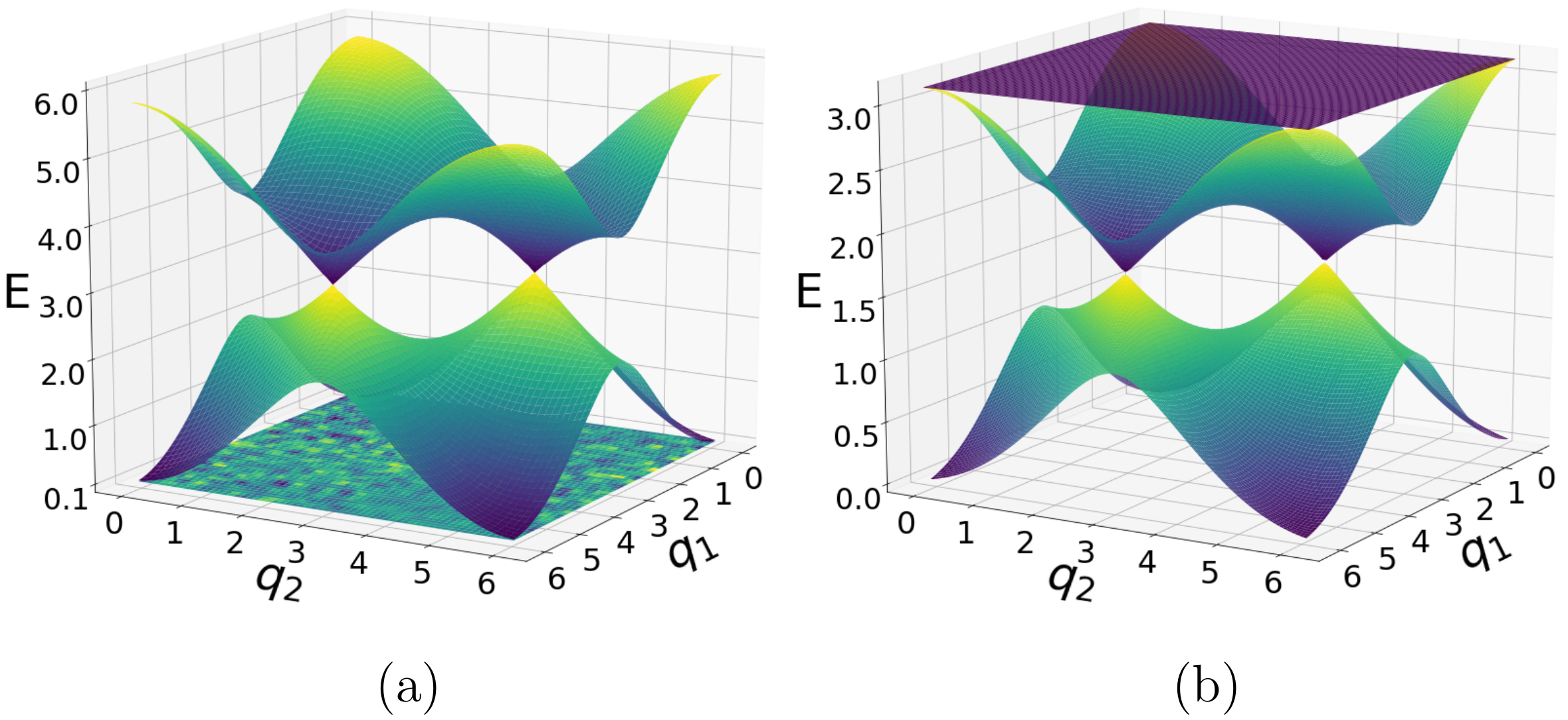}
   \caption{Spectrum of the spin waves in the harmonic approximation in (a) the ferrromagnetic phase $(J_{xy}/J_z=1.9)$ and in (b) the antiferromagnetic phase $(J_{xy}/J_z=2.1$). Energies are defined in units of the $J_z$ coupling for $s=1/2$, and we show the spectrum of the free branches of the spin wave excitations.}
          \label{}
    \vskip -0.5cm
    \end{figure}

To study quantum effects, we analyse the spin wave spectrum in both phases adapting the calculation of Ref. \cite{Kallin} for the present situation. In the ferromagnetic phase, we check the presence of a quadratic dispersion relation in the vicinity of $|{\bf k}|=0$ with an energy of $2sJ_z(2-\gamma_F+\frac{\gamma_F}{2}|{\bf k}|^2)$ where $\gamma_F=J_{xy}/J_z$. This dispersive branch approaches the classical energy when $J_{xy}\sim 2 J_z$ and corresponds to adiabatic deformations of the phase difference $\phi_i-\phi_j$ for nearest neighbors around zero. In addition, we check the presence of a flat band corresponding to alternating 0 and $\pi$ values of the phases 
$\phi_i$ for the six sites forming an honeycomb cell \cite{Kallin}. The flat band energy also meets the classical energy at the phase transition. Taking into account the entropy at finite temperature, corresponding to degenerate states associated to the (free) angles $\phi_i$, then the free energy of this flat band should be lowered compared to the classical ferrromagnetic state when $J_{xy}=2J_z$, justifying that the ferromagnetic ground state is not the correct classical ground state. The spin system rotates in the $xy$ plane forming an antiferromagnetic phase where spin vectors order at $120^o$. 
In the antiferromagnetic phase, the spins lock in the $xy$ plane according to $\phi_i-\phi_j=2\pi/3$ in radians, and the flat band now moves at higher energy as shown in Fig. 3. In this case, the flat band would rather correspond to out-of-plane staggered spin excitations.   The energy of the spin waves for the lowest dispersive band, for $|{\bf k}|\ll 1$, is given by $2J_{xy} s\sqrt{1-2\gamma_{AF}}|{\bf k}|$ with $J_z=\gamma_{AF}J_{xy}$ and $\gamma_{AF}=1/\gamma_F$. The linear dispersion of the spin waves corresponds to adiabatic deformations of $\phi_i-\phi_j$
for nearest neighbors around the value $2\pi/3$ in Eq. (\ref{bloch}). 

We have checked the robustness of our results when including a coupling between two successive Kagome layers \cite{structure}; see Supplemental Material in Ref. \cite{SM}.

{\it Topological Transition.---} Here, we describe the effect of the magnetic transition on the conduction electrons for 2/3 on-site occupancy, first assuming that the fluctuations in the Hund's coupling direction are small. If the amplitude of $h_c$ is sufficiently large compared to $t$, we can write $\langle s_i^z\rangle = \langle S_i^z\rangle$ \cite{Karyn}, where $s_i^z=(c^{\dagger}_{i\uparrow}c_{i\uparrow}-c^{\dagger}_{i\downarrow}c_{i\downarrow})/2$ represents the conduction electron's magnetization. Writing $\langle s_i^z\rangle=(P_{\uparrow}-P_{\downarrow})/2$ with $P_{\uparrow}+P_{\downarrow}=1$,  in that case we predict: $P_{\uparrow}=1$ and $P_{\downarrow}=0$ if $J_{xy}<2J_z$ corresponding to the ferromagnetic order along $z$ direction and $P_{\uparrow}=P_{\downarrow}=1/2$ if $J_{xy}>2J_z$ in the antiferromagnetic phase. When the spins align in the $xy$ plane they do not modify the motion of the itinerant electrons when $\langle S_i^z\rangle=0$ in Eq. (\ref{hc}).  This produces a QAH-QSH transition at the magnetic transition associated with a change of band topology. In the ferromagnetic case, as studied above, the lowest and middle bands associated with spin-up particles are filled, whereas in the QSH phase the lowest band associated to each spin species is completely filled, whereas middle and upper bands are empty (see Fig. 2). The QSH phase occurs because the spin-down particles are described by an opposite phase $-\Phi$ compared to the spin-up particles on a triangle, if we generalize the form of the spin-orbit coupling as in the Kane-Mele model, $i\lambda\sum_{\langle i;j\rangle} \nu_{ij}(c^{\dagger}_{i\uparrow}c_{j\uparrow}-c^{\dagger}_{i\downarrow}c_{j\downarrow})$, which is reminiscent of an atomic spin-orbit coupling $L^z s^z$ with ${\bf L}$ being the angular momentum of electrons on a lattice. The spin-up and spin-down electrons are described by the same nearest-neighbor hopping amplitude in the antiferromagnetic phase such that $t_{\uparrow}=t_{\downarrow}=t$. The core spins act as a local magnetic field which breaks time-reversal symmetry if the net magnetization on a triangle is non-zero. In the antiferromagnetic phase, the sum of the three arrows describing the spins in a triangle is zero, therefore time-reversal symmetry is preserved if $\langle S_i^z\rangle=0$ and a $\mathbb{Z}_2$ topological order can develop, where the Chern number of each lowest band is equal to $C_l^{\uparrow}=-C_l^{\downarrow}=+\hbox{sgn}(\sin\Phi)$. 

For Co$_3$Sn$_2$S$_2$, it is important to emphasize that the ferromagnetic fraction varies smoothly with temperature or here the ratio $J_{xy}/J_z$, which breaks time-reversal symmetry. In our approach, it produces a QAH conductance at the edges which is proportional to $2(e^2/h)\langle s_i^z\rangle(J_{z0})=2(e^2/h)\langle S_i^z\rangle(J_{z0})=(e^2/h)\langle \cos\theta\rangle(J_{z0})$; here, $h$ corresponds to the Planck constant and $e$ is the charge of an electron. Here, we include the effect of fluctuations in the direction of the Hund's coupling. Such fluctuations induce a slightly disordered distribution of $J_z$ parameters that we study globally, with the same mean $J_{z0}$ and with the same variance $\sigma$ at each site. The symbol $\langle ...\rangle(J_{z0})$ refers to an ensemble averaged value, for instance, on different sample realizations.  These variations on the value of $h_c$ could be produced by temperature effects generating a random (noisy) Hund's coupling along $z$ direction. This variance could also be stabilized by a Dyaloshinskii-Moriya term $D_{ij}{\bf S}_i\times{\bf S}_j$ producing weak ferromagnetism along $z$ direction in the antiferromagnetic $120^o$ phase \cite{fiete,Rosales}. 

\begin{figure}[t]
   \includegraphics[width=8cm]{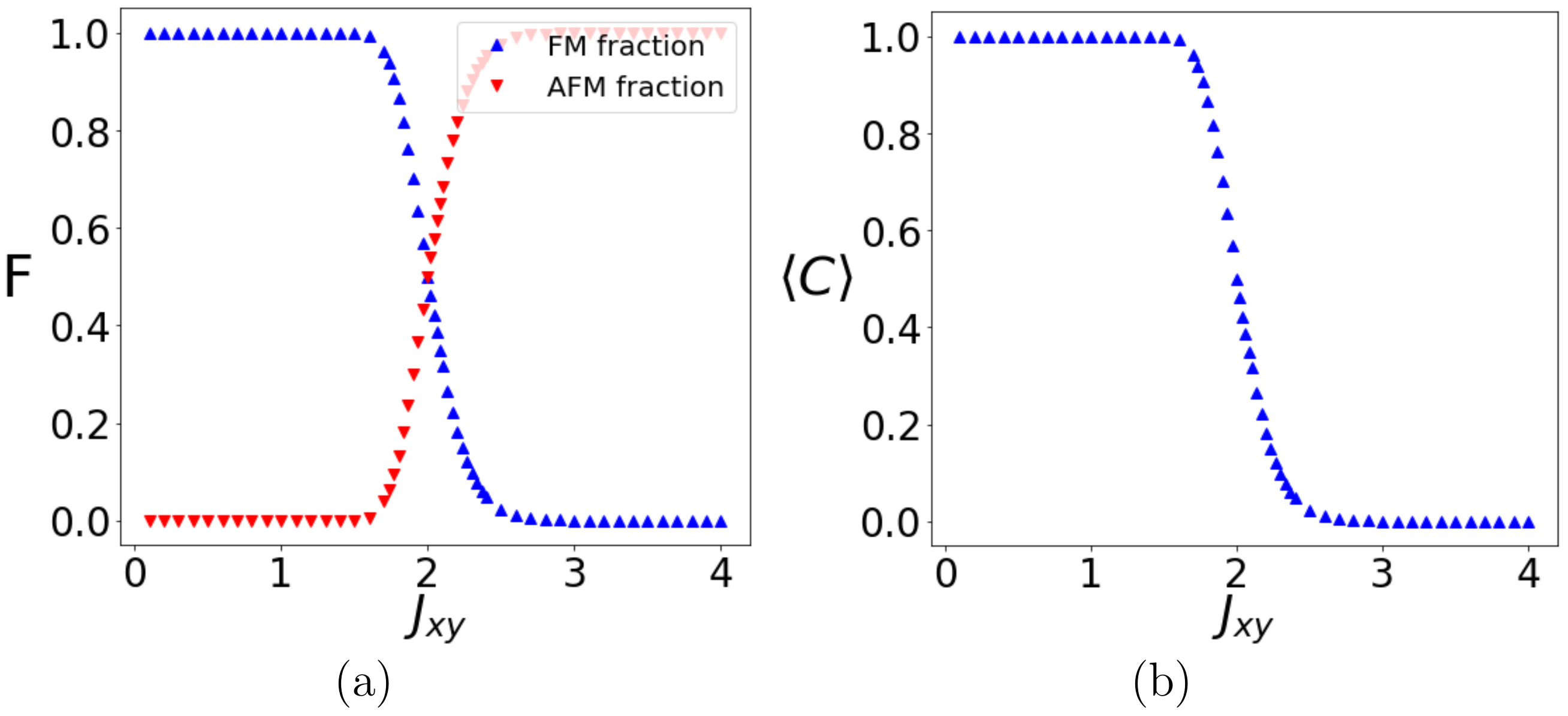}
   \caption{(a) Fractions of the ferromagnetism $F=\langle m_z\rangle(J_{z0})$ and antiferromagnetism (1-F), and (b) averaged Chern number as a function of $J_{xy}/J_{z0}$, taking into account fluctuations in the Hund's coupling. The ratio $\lambda/t$ is set to 0.1 and the standard deviation $\sigma$ of the $J_z$-distribution is set to $0.05J_{xy}$.}
          \label{fraction}
    \vskip -0.5cm
\end{figure}

Then, we study the effect of such fluctuations on bulk properties. We take the distribution of $J_z$ couplings as Gaussian, $P(J_z;J_{z0})=\frac{1}{\sqrt{2\pi}\sigma}e^{-(J_z-J_{z0})^2/2\sigma^2}$ with a variance $\sigma\ll J_{z0}$ (or much smaller than $k_B$T for Co$_3$Sn$_2$S$_2$). For $2J_{z0}=J_{xy}$, now the system can show a coexistence between ferromagnetism along $z$ direction and antiferromagnetism in the $xy$ plane. Introducing $m_z=2\langle S_i^z \rangle=\cos \theta$, then $m_z=+1$ if $J_{xy}<2J_z$ and $m_z=0$ if $J_{xy}>2J_z$. The ensemble average value of $m_z$ including the Gaussian fluctuations is given by:
\begin{eqnarray}
\langle m_z\rangle(J_{z0}) &=& \int_0^{+\infty} dJ_z P(J_z;J_{z0})m_z(J_z) \\ \nonumber
&=& \frac{1}{2}\hbox{erfc}\left[\frac{1}{\sqrt{2}\sigma}\left(\frac{J_{xy}}{2}-J_{z0}\right)\right],
\end{eqnarray}
where erfc corresponds to the complementary error function. 
For the conduction electrons, if $J_{xy}<2J_z$ we have a (sample with a) Chern number $C=C_l^{\uparrow}=+1$ corresponding to $P_{\uparrow}=1$ and $P_{\downarrow}=0$ and for $J_{xy}>2J_z$ we have $C=C_l^{\uparrow}-C_l^{\downarrow}=0$ corresponding to $P_{\uparrow}=P_{\downarrow}=1/2$. Therefore, we introduce the averaged Chern number 
\begin{equation}
\label{Cmz}
\langle C \rangle(J_{z0}) = \int_0^{+\infty} dJ_z P(J_z;J_{z0}) C(J_z) = \langle m_z\rangle(J_{z0}).
\end{equation}
 In Fig. \ref{fraction}, we show the behavior of the averaged Chern number and averaged magnetization. Eq. (\ref{Cmz}) relates the progressive evolution of the magnetization along z-axis in the bulk with the (averaged) Chern number, as observed in Refs. \cite{Princeton,Dresden}.  We reproduce a  bulk-edge correspondence where the conductance at the edges takes the form $(e^2/h)\langle C\rangle(J_{z0})$. In Fig. \ref{fraction}, we draw the evolution of the Chern number for 2/3 on-site occupancy for the itinerant electrons. A transition from QSH to QAH effect was also reported in HgTe materials when doping with random magnetic Mn dopants \cite{Wurzburg}, and in thin films of (Bi,Sb)$_2$Te$_3$ doped with Cr-atoms \cite{Chang}. 

To summarize, we have built a model taking into account both localized electrons giving rise to a magnetic transition and conduction electrons producing topology of Bloch bands on the Kagome lattice. We hope that this may participate to the understanding of the quantum material Co$_3$Sn$_2$S$_2$ and a similar theoretical approach could be developed to describe  Fe$_3$Sn$_2$ Kagome bilayer systems \cite{Aeppli}. Changing the stochiometry of a Co-atom Kagome plane, open questions remain including the precise value of the Chern number \cite{Lukas}.
\\

We acknowledge discussions with Joel Hutchinson, Philipp Klein, Alexandru Petrescu, Nicolas Regnault and Jakob Reichel. This work is funded through ANR BOCA and the Deutsche Forschungsgemeinschaft via DFG 
FOR2414 under Project No. 277974659. KLH also acknowledges the Canadian CIFAR for support. 

\onecolumngrid

\begin{center}
{\bf Chern number calculation, edge modes, lattice currents and local density of states}
\end{center}

{\it Chern number and Hamiltonian.---}
The Chern number is evaluated from the formula (see \cite{Petrescu})
\begin{equation}
C(E_F) = \dfrac{1}{2 \pi} \sum_{n,\sigma} \int_{\textrm{BZ}} d^2\boldsymbol{k} \, \Theta(E_F-E_{n,\sigma}(\boldsymbol{k})) \, \partial_{\boldsymbol{k}} \times \mathcal{A}_{n,\sigma}(\boldsymbol{k}).
\end{equation} 
The integral is taken over the Brillouin zone and the summation takes into account all the energy bands labelled by ${n=\{1,2,3\}}$ and spin polarization $\sigma=\{\uparrow,\downarrow\}$. It is restricted to the states with energy inferior to the Fermi energy $E_F$. ${\mathcal{A}_{n,\sigma}(\boldsymbol{k}) = - i \bra{n,\sigma,\boldsymbol{k}} \partial_{\boldsymbol{k}} \ket{n,\sigma,\boldsymbol{k}}}$ is the Berry gauge field with $\ket{n,\sigma,\boldsymbol{k}}$ the energy eigenstate associated to the $n^{th}$ energy band $E_{n,\sigma}(\boldsymbol{k})$ of the spin $\sigma$ electron species. The Fermi energy $E_F$ is determined by the number of electrons $N_e=(N_{\uparrow}+N_{\downarrow})=N_e(P_{\uparrow}+P_{\downarrow})$, where we take into account the magnetism from localized spins. Referring to the classical analysis on the magnetism in the Letter, we have the relations
$\langle s_i^z\rangle = (P_{\uparrow} - P_{\downarrow})/2 = \langle S_i^z\rangle$ with $P_{\uparrow}+P_{\downarrow}=1$, such that $P_{\uparrow}=(1/2)(1+\cos\theta)$ and $P_{\downarrow}=(1/2)(1-\cos\theta)$. In the ferromagnetic phase, $P_{\uparrow}=1$ and $P_{\downarrow}=0$ and for the antiferromagnetic phase, if $\langle S_i^z\rangle=0$, then $P_{\uparrow}=P_{\downarrow}=1/2$. 

The effect of the strong Hund's coupling is tackled through the on-site occupancies $P_{\uparrow}$ and $P_{\downarrow}$. This is equivalent to include the effect of a sufficiently strong Hund's coupling $-h_c\sum_i\langle S_i^z\rangle s_i^z$ in the Hamiltonian
written in the wave-vector basis such that $\langle s_i^z\rangle=\langle S_i^z\rangle$, and this has the advantage of writing a common Fermi energy for the total system. 

\begin{figure}[t]
   \includegraphics[width=18cm]{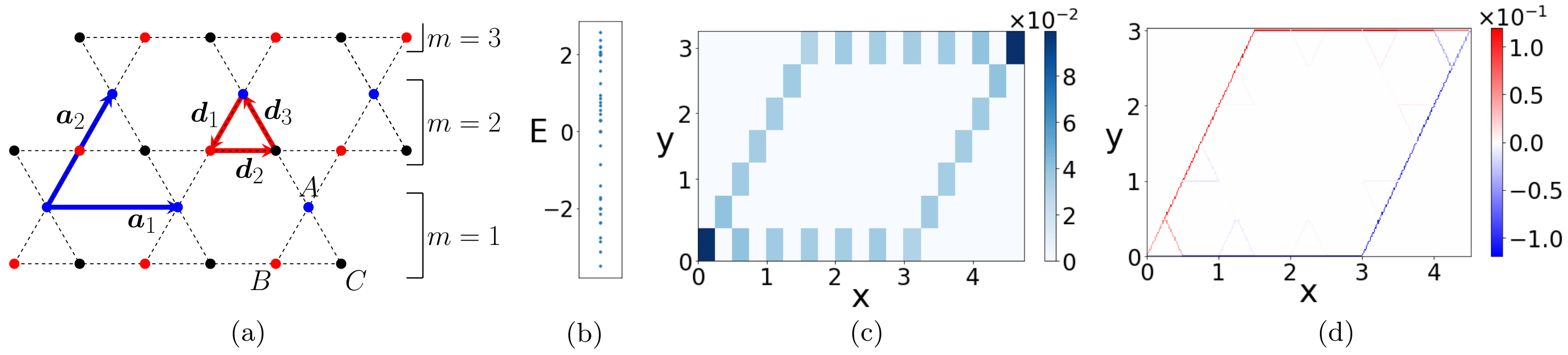}
   \caption{(a) Cylinder geometry on the kagome lattice. The unit cells in the open direction are labeled by integer m. (b) Energies (in units of t) of the plane system composed of 9 unit cells (total of 40 atoms) with flux inside each triangle of the lattice $\Phi=3\pi/4$, (c) local density of states at $E \approx 1.25t$ when the bulk is insulating (corresponding to the $28^{th}$ eigenenergy) and (d) current expectation value over the lattice for the spin up electron species (in units of $t/|\boldsymbol{a}_1|$). The color blue or red refers to the vector associated with the current orientation along the edge. If we take an horizontal path, blue means ``going left" and red ``going right". For another path, if we go up along this path, the color then is ``red" and if we go down the color is ``blue".}
          \label{cylinder}
    \vskip -0.5cm
\end{figure}

Writing  $\Psi_{\boldsymbol{k},\sigma}^\dagger = \left(A_{\boldsymbol{k},\sigma}^\dagger, B_{\boldsymbol{k},\sigma}^\dagger, C_{\boldsymbol{k},\sigma}^\dagger\right),$ with $A_{\boldsymbol{k},\sigma}^\dagger, B_{\boldsymbol{k},\sigma}^\dagger$ and $C_{\boldsymbol{k},\sigma}^\dagger$ the creation operators for the itinerant spins on respectively atoms $A$, $B$ and $C$ sketched in Fig.~\ref{cylinder}(a), then the momentum space Hamiltonian reads ${H = \sum_{\sigma=\{\uparrow,\downarrow\}} \sum_{\boldsymbol{k}} \Psi_{\boldsymbol{k},\sigma}^\dagger h_{\boldsymbol{k},\sigma} \Psi_{\boldsymbol{k},\sigma},}$ with
\begin{equation} 
    h_{\boldsymbol{k},\sigma} = -2 r
\begin{pmatrix}
\mu_\sigma h_a/4 r & e^{-i\Phi_\sigma/3} \cos (\boldsymbol{k}.\boldsymbol{d}_1) & e^{i\Phi_\sigma/3} \cos (\boldsymbol{k}.\boldsymbol{d}_3) \\
e^{i\Phi_\sigma/3} \cos (\boldsymbol{k}.\boldsymbol{d}_1) & \mu_\sigma h_a/4 r &e^{-i\Phi_\sigma/3} \cos (\boldsymbol{k}.\boldsymbol{d}_2) \\
e^{-i\Phi_\sigma/3} \cos (\boldsymbol{k}.\boldsymbol{d}_3) & e^{i\Phi_\sigma/3} \cos (\boldsymbol{k}.\boldsymbol{d}_2) & \mu_\sigma h_a/4 r
\end{pmatrix},
\end{equation}
and  $-r e^{i\Phi_\uparrow/3}\equiv -(t + i\lambda) $, $\Phi_\uparrow = \Phi = -\Phi_\downarrow$ and $\mu_\uparrow=1=-\mu_\downarrow$. Displacement vectors $\boldsymbol{d}_1$, $\boldsymbol{d}_2$ and $\boldsymbol{d}_3$ are sketched in Fig.~\ref{cylinder}. We numerically find the Chern number applying a gauge independent formula for the Berry curvature $\partial_{\boldsymbol{k}} \times \mathcal{A}_{n,\sigma}(\boldsymbol{k})$ (see \cite{Fukui}) and checked the agreement with the analytical prediction \cite{KagomeNagaosa}. \newline

{\it Edge modes, lattice currents and LDOS.---} Here we evaluate the eigenenergies $E_\Psi$ and the eigenstates $\Psi$ of the Hamiltonian, for a lattice cylinder geometry with periodicity along direction $\boldsymbol{a}_1$ (see Fig. \ref{cylinder}(a)), as in Ref. \cite{Petrescu}. We also compute it along with the currents for a plane geometry composed of 9 unit cells. We consider the $h_a \rightarrow 0$ limit and we take into account the Hund's coupling as a constraint on $N_\uparrow$ and $N_\downarrow$, as described above. We write the Hamiltonian $H=\sum_\sigma H_\sigma$ and, for the cylinder geometry, we apply a partial Fourier transform on $H_\sigma$. For each spin $\sigma$ species, a numerical computation gives the eigenenergies and the local density of states given by $\rho_\sigma(E,\boldsymbol{r})\equiv \sum_{\Psi_\sigma} \delta(E-E_\Psi) |\langle {\boldsymbol{r}}|{\Psi_\sigma}\rangle|^2$. It satisfies the following normalization condition : $\int dE \sum_{\boldsymbol{r} \in \boldsymbol{R}} \rho_\sigma(E,\boldsymbol{r})=N_a$, with $\boldsymbol{R}$ the ensemble of all possible atomic sites and $N_a$ their total number.

For the cylinder geometry, at $E=+1.5t$ for instance, we observe edge modes. The energy and the expression of these edge modes can be analytically determined. We write the eigenstates as a superposition of localized states on each atom : $\Psi_\sigma(q_1) \equiv \sum_{\alpha = A,B,C} \sum_m \phi_{\alpha,m,\sigma}(q_1) \ket{\alpha, m, \sigma} $. From the numerical evaluation, we assume that $\phi_{A,m}(q_1) = 0$ and we use the ansatz $\phi_{B,C}(m) = \lambda_\sigma^m \phi_{B,C}(0).$
It gives the energies of the 2 edge modes $E_\pm$ and the associated values of $\lambda_{\pm,\sigma}$, in accordance with the analytical solution:
$$E_\pm= \pm 2r \cos (\dfrac{q_1}{2}), \quad \lambda_{+,\sigma}=-  \left. \cos \left( \dfrac{\Phi_\sigma}{2}-\dfrac{q_1}{4}\right)\right/ \cos \left( \dfrac{\Phi_\sigma}{2}+\dfrac{q_1}{4}\right), \quad \lambda_{-,\sigma}= \left. \sin \left( \dfrac{\Phi_\sigma}{2}-\dfrac{q_1}{4}\right)\right/ \sin \left( \dfrac{\Phi_\sigma}{2}+\dfrac{q_1}{4}\right).$$
Whether $|\lambda_{\pm,\sigma}|>1$ or  $|\lambda_{\pm,\sigma}|<1$ determines at which edge of the system the mode is localized. We notice that $\lambda_{\pm,\sigma}(q_1)=\left. 1\right/\lambda_{\pm,\sigma}(-q_1)$ and that $\lambda_{\pm,\sigma}(q_1)=\left. 1\right/\lambda_{\pm,\overline{\sigma} }(q_1)$ with $\overline{\sigma}=\downarrow$ ($\uparrow$) if $\sigma=\uparrow$ ($\downarrow$). This and the fact that the energy of each eigenmode is a cosine centered around $q_1=0$ predicts one eigenmode located at each edge of the system in the ferromagnetic phase and two counter-propagating eigenmodes at each edge of the system in the antiferromagnetic phase. 

For the plane geometry we also observe edge modes for certain values of the Fermi energy (see Figs. 1(b) and 1(c)). The currents operator between two lattice sites $m$ and $n$ is given by $j_{mn}\equiv -i a_m^\dagger t_{mn} a_n + H.c.$. Over the lattice, the current expectation value $\sum_{\Psi_\sigma} \delta(E-E_\Psi) \bra{\Psi_\sigma}j_{mn}\ket{\Psi_\sigma}$ is localized on the edges for certain values of the Fermi energy (sketched in Fig. 1(d) for one spin species) and the one associated to the spin up electrons are opposite to the one associated to the spin down electrons.

\begin{center}
{\bf Interplane coupling}
\end{center}

{\it Double-layer model.---} Let us discuss a double-layer model in order to show that the results of our study can be generalized to a 3-dimensional model. The 3-dimensional structure of Co$_3$Sn$_2$S$_2$ shows covalent bonds between the different planes, forming stripes of alternating Co and Sn atoms (see fig. 4b of \cite{structure}). Therefore it is rather intuitive to consider the following term for the inter-plane coupling part of the effective Hamiltonian describing the so-called local spins
\begin{equation}
\widetilde{H}_{S} = \sum_{\langle \langle i;j\rangle \rangle} [-\widetilde{J}_z S_i^z S_j^z - \widetilde{J}_{xy}(S_i^xS_j^x + S_i^y S_j^y)],
\end{equation} 
where i and j are inter-plane nearest neighbors which make part of the same stripe of Co (and Sn) atoms. Adding this term $\widetilde{H}_S$ to the in-plane magnetic contribution, we easily see that the local spins carried by the Co inter-plane stripe show a ferromagnetic or antiferromagnetic order depending on the (inter-plane) coupling parameters sign. In both cases, this stabilizes the plane phases we described in our study.

\bibliography{main.bib}

\end{document}